\documentclass[aip,reprint]{revtex4-1}
\usepackage{graphics,amsmath,color}

\begin{document}

\title{Photoluminescence from voids created by femtosecond laser pulses inside cubic-BN} 



\author{R. Buividas}

\affiliation{Centre for Micro-Photonics, Swinburne University of
Technology, John st., Hawthorn, Vic 3122, Australia}
\author{I. Aharonovich}
\affiliation{School of Mathematical and Physical Sciences,
University of Technology Sydney, Thomas St, Ultimo, NSW 2007
Australia}
\author{G. Seniutinas}
\affiliation{Centre for Micro-Photonics, Swinburne University of
Technology, John st., Hawthorn, Vic 3122, Australia}
\author{X. W. Wang}
\affiliation{Centre for Micro-Photonics, Swinburne University of
Technology, John st., Hawthorn, Vic 3122, Australia}
\author{L. Rapp}
\affiliation{Laser Physics Center, Research School of Physics {\&}
Engineering, Australian National University, ACT 0200, Australia}
\author{A. V. Rode}
\affiliation{Laser Physics Center, Research School of Physics {\&}
Engineering, Australian National University, ACT 0200, Australia}
\author{T. Taniguchi}
\affiliation{National Institute for Materials Science, 1-1 Namiki
Tsukuba Ibaraki 305-0044 Japan}
\author{S. Juodkazis}
\email[]{SJuodkazis@swin.edu.au}
\affiliation{Centre for Micro-Photonics, Swinburne University of
Technology, John st., Hawthorn, Vic 3122, Australia}
\date{\today}
\begin{abstract}
Photoluminescence (PL) from femtosecond laser modified regions
inside cubic-boron nitride (c-BN) was measured under UV and
visible light excitation. Bright PL at the red spectral range was
observed, with a typical excited state lifetime of $\sim 4$~ns.
Sharp emission lines are consistent with PL of intrinsic vibronic
defects linked to the nitrogen vacancy formation (via Frenkel
pair) observed earlier in high energy electron irradiated and
ion-implanted c-BN. These, formerly known as the radiation
centers, RC1, RC2, and RC3 have been identified at the locus of
the voids formed by single fs-laser pulse. The method is promising
to engineer color centers in c-BN for photonic applications.
\end{abstract}

\maketitle

Investigation of high pressure and temperature phases of different
materials are spanning from the fundamental
research~\cite{Gleason,Fletcher,Xu} as a recently demonstrated
metallic bonding in hydrogen~\cite{Knudson} to developing new
pathways of material synthesis for future practical
use~\cite{McMillan}. Structural modifications inside transparent
materials can be controlled down to sub-wavelength
resolution~\cite{Shimotsuma2003,Hnatovsky} by direct laser write
and opens possibility to create micro-optical~\cite{Beresna2011a}
and mechanical~\cite{Bellouard2005} elements,
waveguides~\cite{Burghoff2007,Ams2009,DellaValle2009},  to render
a chemically resistant material wet-bath etchable for microfluidic
applications~\cite{01ol277}. Of a particular interest are wide
band gap materials, such as Cubic BN (c-BN) with a bulk modulus of
$\sim 400$~GPa and a wide bandgap of 6.5~eV. It is made by the
diamond anvil cell (DAC) or belt-type high pressure compression of
a hexagonal phase h-BN~\cite{Taniguchi}.

Optical properties of boron nitrides are attracting an increasing
interest due to their ability to host optically stable single
photon emitters at room temperature~\cite{Aharonovich1} and their
hyperbolic properties. To this extent, sub-bandgap excitation of
controllably formed and patterned defects in c-BN is strongly
anticipated. Moreover, theoretical modeling predicts an analog of
the nitrogen-vacancy NV$^-$ centers known in diamond for the
B-vacancy oxygen pair in c-BN~\cite{Abtew}, which is yet to be
confirmed experimentally. The stacking fault energy $191\pm
15$~mJ.m$^{-2}$ in c-BN is comparable to that in
diamond~\cite{Nistor} and a similar defect formation is expected.

\begin{figure}[tb]
\resizebox{0.5\textwidth}{!}{\includegraphics{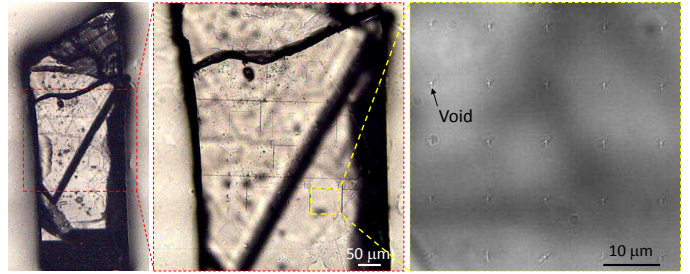}}
\caption{Sample of c-BN crystal and close up views of the laser
structured regions. The left image shows array fabricated by $E_p
= 27$~nJ energy pulses at 515~nm wavelength focused with an
objective lens of numerical aperture $NA = 1.42$ at depth of $\sim
10~\mu$m; pulse duration was 230~fs.} \label{f-sam}\end{figure}

Here, we employ a femtosecond (fs)-laser irradiation
technique~\cite{10jpd145501,99prb9959} to create optically active
color centers in c-BN. Using confocal microscopy and time resolved
measurements, we show that the formed voids are extremely bright
PL sources and originates from the N-vacancy related radiation
center RC-defects~\cite{Shishonok} observed earlier in c-BN only
by high energy particle or ionising radiation
exposure~\cite{Erasmus}.

Samples of c-BN were grown on a belt-type high pressure equipment~\cite{Taniguchi} and were used for experiments of void
formation by single femtosecond (fs-)laser pulses tens of
micrometers below the surface (Fig.~\ref{f-sam}).

\begin{figure}[tb]
\resizebox{0.4\textwidth}{!}{\includegraphics{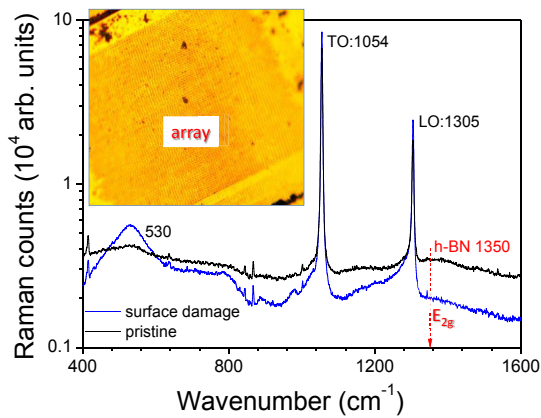}}
\caption{Raman spectra of surface ablation array recorded with
800~nm/150~fs pulses. Excitation wavelength of Raman scattering
was 785~nm. Inset shows field of view approximately $0.1\times
0.1$~mm$^2$.} \label{f-rama}\end{figure}

Laser pulses of $\tau_p = 230$~fs duration at fundamental $\lambda
= 1030$~nm and second harmonic 515~nm wavelengths were focused at
a $10-20~\mu$m depth below the surface of a facet plane inside
c-BN crystals (Fig.~\ref{f-sam}). Tight focusing with numerical
aperture $NA = 1.42$ was implemented to form arrays of damage
sites recorded at different pulse energies $E_p$ at a single pulse
conditions. Separation between irradiation spots was 10~$\mu$m to
eliminate a cross talk for void-formation and optical
characterization. For comparison, dense array of ablation patterns
were fabricated with 800~nm/150~fs pulses focused with $NA = 0.95$
on the surface of c-BN. Judgement of the void presence at the
center of irradiated spot inside c-BN was made by a sharp optical
contrast change which was identical to the void formation in
crystalline sapphire, quartz and glasses of different refractive
indices~\cite{06prl166101,06apa337}; focused ion milling will be
implemented next to reveal an internal structure of the void as it
was made for voids at the Si-SiO$_2$ interface~\cite{Andrei}.

Photoluminescence and its transients were measured under laser
diode 405~nm/30~ps or 510~nm/100~ps (PiLAS; advanced Laser Diode
systems) excitation using $NA=0.7$ objective lens and a single
photon counting avalanche photo diode (SPCM-AQRH-14) as a
detector. A spectral window of PL collection was filter selected
at 620-650~nm. A piezo-scanner was implemented to record a PL map
around the void-structures. PL spectra were recorded by a
spectrometer (Princeton Instruments). Raman spectra were acquired
with an InVia Streamline microscope (Renishaw, UK) under 785~nm
excitation and $NA = 0.4$ focusing.

\begin{figure}[tb]
\resizebox{0.4\textwidth}{!}{\includegraphics{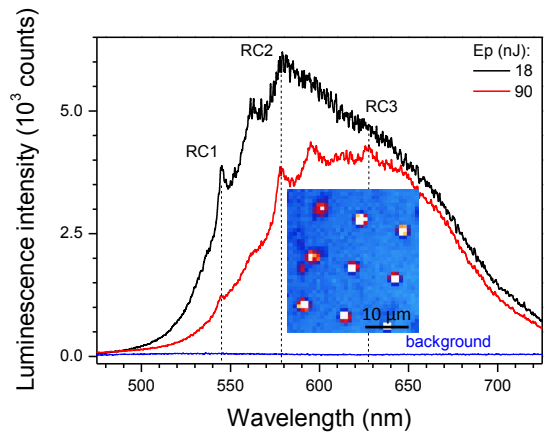}}
\caption{Photoluminescence (PL) spectra from void-structures in
c-BN. PL excitation wavelength was 405~nm. The inset: a PL map
under 532~nm excitation; the void-structures were made with $E_p =
18$~nJ single pulses of 515~nm wavelength at a 10~$\mu$m depth.}
\label{f-405}
\end{figure}

Strong optical contrast changes were observed at the focus of
tightly focused $NA = 1.42$ fs-laser pulse in c-BN at the
threshold values of $E_p = 4.5$~nJ ($\lambda = 515$~nm) and 80~nJ
(1030 nm) estimated.  Due to high refractive index of $n \simeq
2.1$~\cite{Cappellini} at the used wavelength, a spherical
aberration defines the focal volume which became slightly
larger~\cite{03apa257} as compared with the diffraction-limited
focal size of diameter $d = 1.22\lambda/NA$. At such tight
focusing, a self focusing is avoided since 10~nJ pulse corresponds
to only 43~kW/pulse power but reach 7.1~TW/cm$^2$ irradiance for
$\lambda = 1030$~nm. These are direct write conditions with
irradiance of $\sim 10$~TW/cm$^2$ when voids are formed in
different transparent materials under equivalent
focusing~\cite{06prl166101,06apa337}. The voids are of
sub-wavelength $50-200$~nm in
diameters~\cite{06prl166101,06apa337} and their actual size has to
be measured using focused ion beam cross sections. Micro-cracks
comparable in size with the central void-structure were observed
in c-BN when the void-structure was created by a single focused
fs-laser pulse (Fig.~\ref{f-sam}). These strongly modified regions
were subject of further examination using PL (Fig.~\ref{f-405}).

When void is formed in c-BN it could be expected that relaxation
into a less dense phase of h-BN or into an amorphous phase occurs.
To test this conjecture an ablation on the surface was carried out
and Raman scattering measured. Characteristic transversal and
longitudinal optical phonon modes TO and LO, respectively, were
observed with small broadening at the high-energy side of the
modes. Back-scattered Raman signal excited by 785~nm irradiation
was collected with a $NA = 0.4$ lens averaging response from area
with several ablation sites made by single pulses. Wide peak at
520~cm$^{-1}$ (65.7~meV) was the strongest modification observed
from surface ablated regions made by 800~nm/150~fs pulses.
Presence of a lower density h-BN was not confirmed from ablated
regions which would be recognisable by its 1350~cm$^{-1}$ E$_{2g}$
mode~\cite{Reich}. Widening of the Raman peak at the LO and TO
modes could be related to disordering at the void region, however
there is no obvious shift to smaller wavenumbers which takes place
for nano-crystallites of c-BN~\cite{Werninghaus}.

\begin{figure}[tb]
\resizebox{0.5\textwidth}{!}{\centering\includegraphics{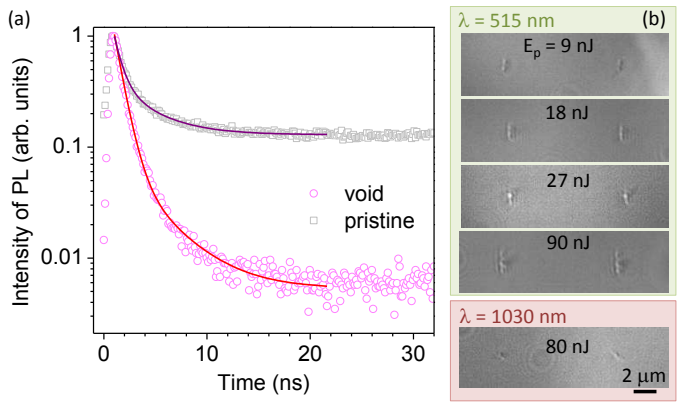}}
\caption{(a) Photoluminescence (PL) transients from the void
structure and from undamaged c-BN excitated by 510~nm/100~ps
illuminaiton using objective lens $NA = 0.7$. Void was formed by
$E_p = 27$~nJ single pulses of 515~nm wavelength at a 10~$\mu$m
depth. (b) Optical images of the typical laser damaged sites with
void present at the center for the brightest central spot;
void-structures made at different pulse energies $E_p$ and 515 and
1030~nm wavelength.} \label{f-decay}
\end{figure}

Confocal microscopy with 405~nm excitation source was employed to
characterize the fabricated voids-structures in c-BN.
Photoluminescence of nitrogen is well studied in atmospheric
discharge experiments and lightning observations with lines at
deep-UV and at 455, 556, 577~nm identified as atomic neutral
nitrogen N~\cite{Uman}. Under 405~nm illumination, PL from the
void regions has recognizable features within similar spectral
range (Fig.~\ref{f-405}). Molecular N$_2$ PL which occurs in
300-400 nm window was out of the range of observation in this
first experiment. The confocal map is shown in inset of
Fig.~\ref{f-405}. The bright spots correspond to the location of
the voids.

The observed PL features are matching perfectly the intrinsic
vibronic defects RC1,2,3 reported in c-BN irradiated by high
energy 1.9~MeV electrons~\cite{Shishonok,Erasmus}. It is
noteworthy that PL was measured at room temperature and no
post-irradiation annealing was required as it is usually the case
after ion implantation. The intrinsic Frenkel pair defects
due to the N-vacancy formation earlier observed in
cathodo-luminescence (CL) and identified as RC1 with zero phonon
line of (2.27~eV, 546.2~nm), RC2 (2.15~eV, 576.7~nm) and RC3
(1.99~eV, 623.1~nm)~\cite{Shishonok,Erasmus} were found matching
very well the defect PL we observed from  the voids made by
fs-laser pulses (see, Table~\ref{t1}). Deterministic fabrication of these color centesr is demonstrated here for the first time using ultra short light pulses.

\begin{table}
  \centering\caption{Emission of RC centers created by electron and photon irradiation.}
\begin{tabular}{|c|c|c|}
  \hline
  Defect & CL [nm] & PL [nm] \\
  Radiation       & 4.5~MeV electrons~\cite{Erasmus}& fs-laser pulse\\
  Center (RC)& &(this study)\\
  \hline
  RC1 & 546.2 & $542\pm 2$ \\
  RC2 & 576.7 & $578\pm 2$ \\
  RC3 & 623.1 & $628\pm 2$ \\
  \hline
 \end{tabular}
  \label{t1}
\end{table}

The PL excited at 510~nm/100~ps illumination showed close to a
single exponential decay with time constant of 3.7~ns
(Fig.~\ref{f-decay}(a)), while under the 405~nm/30~ps excitation
the decay was also similar 4~ns (not shown). Long stretched
exponential decay is usually indicative of a recombination of the
electrons and holes trapped on defects which are distributed in
energy and separated spatially by varying distances. Very long
multi-exponential decays were observed in silica glass with
nano-gratings formed by fs-laser irradiation as measured by a
time-domain method~\cite{15pr}. Very similar temporal transients
from pristine regions and void-structures are consistent with
self-trapping of electron-hole pairs (a pathway of excitonic
decay) which is typical for wide bandgap materials. Such scenario
is also consistent with stretched exponential decay and is
observed in pristine regions of crystals and dielectrics.

The RC2 and RC3 centers were observed in CL under 3.5~GPa
pressure~\cite{Shishonok}. Following earlier studies of
void-structures in sapphire and silica~\cite{11nc445,Andrei} even
higher residual pressure were observed at the void region. Future
studies are required to reveal internal morphology of the voids in
c-BN, presence of amorphisation with better spatial resolution.
Small changes of the Raman TO and LO phonon modes from ablated
regions as well as $\pm 5$~nm changes in PL of the RC-defects
might de related to presence of shock amorphised c-BN as it was
observed in sapphire~\cite{06am1361} where voids had shell of
metastable amorphous phase.

Void-formation in c-BN by single fs-laser pulses was observed by
strong optical contrast changes at tens - of - micrometers below
the surface. Three vibronic radiation center defects RC1,2,3 were
identified in c-BN by photon irradiation. PL from single voids
showed fast $\sim 4$~ns transient with stretched exponential
slower tail of the decay. Photoluminescence transients from
regions with defects and without were fitted by the same time
constants. This corroborates an intrinsic character of the Frenkel
pair defects as they are, most probably, formed from the same
pre-cursors which active in self-trapping of photo-carriers in
pristine c-BN. There were no indication of h-BN formation on the
surface nor in the bulk from laser treated volume.  Laser
patterning of defects in wide bandgap materials with
sub-wavelength precision and a 3D capability of their patterning
is appealing for engineering of deterministic sources for
photonics in BN systems.

\begin{acknowledgments}
\small{SJ is grateful for partial support via the Australian
Research Council Discovery project DP130101205 and fs-laser
fabrication setup via a technology transfer project with Altechna
Ltd. Authors are grateful to P. R. Stoddart for access to Raman
setup. Partial support to this work by Air Force Office of
Scientific Research, USA (FA9550-12-1-0482) is gratefully
acknowledged.}
\end{acknowledgments}

\small

\end{document}